# A modification of Oersted experiment

**Dimitar G. STOYANOV**

Sliven Engineering and Pedagogical Faculty, Sofia Technical University
59 Burgasko Shosse Blvd, 8800 Sliven, BULGARIA

E-mail: **dgstoyanov@abv.bg**

**Abstract** The paper describes a simple setup of Oersted experiment. A planar coil of wires has been used to deflect vigorously the magnetic needle (more than 80 angular degrees) when a current of up to 1 A flows along it. Based on theoretical analysis the torque on the magnetic field is analytically expressed taking into account the inhomogeneity of the field and the needle shape. What is more, a procedure to measure the Earth's magnetic component is implied and implemented and its magnitude has been estimated following the same steps..

## 1. Introduction
In 1820 Christian Oersted noticed that a compass needle deflected its initially aligned north-south direction in the presence of a current-carrying wire. That experiment has been the first one to indicate that current-carrying wire produces a magnetic field. However, the classical setup employs a source of steady power supply and a current of 10 – 20 A, which makes the demonstration very difficult.
   Hence, we propose a low-cost type of apparatus as a simple solution and practical modification of Oersted experiment.

## 2. Theoretical Analysis
### 2.1 A Magnetic field of a finite wire
Let a straight finite wire of length $L$ lies on **y**-axis of the Cartesian coordinate system (Fig. 1). The wire is oriented symmetrically to the origin of the coordinate system. A steady electric current, **I**, directed to the positive **y** - axis flows along the wire. The current carried by the wire creates a magnetic field around the wire.

At any point on **z** - axis, at a distance of **z** from the origin of the coordinate system the vector of the magnetic induction, $\vec{B}$, is oriented to the positive direction of **x** - axis (in the setup perpendicularly in the drawing in figure 1, facing the paper).

The magnitude of the magnetic induction vector is obtained by using Biot – Savart law [1-3]:

$$B_x = \frac{\mu_0 . I}{4.\pi} \cdot \int_{-L/2}^{+L/2} \frac{z.dy}{(y^2 + z^2)^{3/2}} \qquad (1)$$

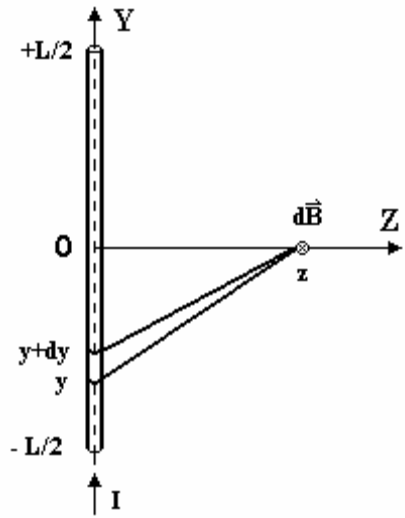

**Figure 1. A magnetic field of a finite current-carrying wire**

The solution to the integral is in [3]:

$$B_x = \frac{\mu_0 . I}{2.\pi.z} \cdot \frac{1}{\sqrt{1 + \left(\frac{2.z}{L}\right)^2}} \cdot \qquad (2)$$

The first multiplier on the right of (2) is the field of a straight infinite wire [1]:

$$B_x(\infty) = \frac{\mu_0 . I}{2.\pi.z}. \qquad (3)$$

The second multiplier is a correction for the (3) final length of the wire. The same multiplier is a continuous function of **L / z** to the value of **1** at **L / z → ∞**. At **L / z = 10**, the multiplier has a value of **0.9806** while at **L / z = 20**, it has a value of **0.9950**.

Hence, at **L / z ≥ 20**, the field created by the wire with an accuracy of 0.5 % is equal to that of an infinite wire. For instance, if **z = 1 cm**, we can use a wire length of **20 cm**.

## 2.2 A magnetic needle in a magnetic field of wire

We would like to point out that the theoretical analyses is totally based on the interaction of the magnetic needle with the magnetic field of a straight infinite horizontal current-carrying wire.

Let a straight infinite metal current-carrying wire lies on y - axis of the Cartesian coordinate system **K** (Fig. 2). A steady current flows along the wire with a magnitude of I, having the same positive direction as **y-axis**.

At a point on **z-axis** the suspension point lies at a distance of **z** from the origin of the coordinate system of a magnetic needle of a length $\Delta$. The pivot of the magnetic needle coincides with **z-axis** while the magnetic needle lies and moves on a plane parallel to **XOY** plane. The actual position of the magnetic needle is characterized by the rotation angle of the needle $\theta$ with respect to the positive direction of **y-axis**. The magnetic needle has a magnetic dipole moment of **m**.

The flowing current creates a magnetic field around the wire. The magnitude of the horizontal component $\mathbf{B_I}$ of the magnetic induction vector in the positive direction of **x-axis** can be obtained by the Biot – Savart law [1]:

$$\mathbf{B_I} = \frac{\mu_0 . I}{2.\pi} \cdot \frac{z}{\rho^2} \tag{4}$$

where:

$$\rho = \sqrt{z^2 + x^2}. \tag{5}$$

Is the distance from the given point with coordinates of (**x, y, z**) to the wire axis.

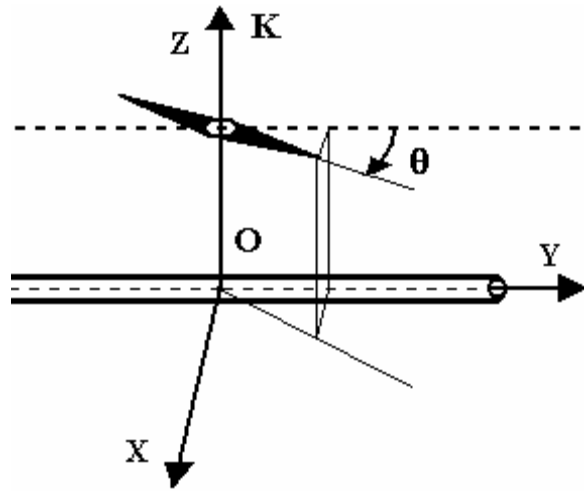

**Figure 2. Mutual disposition of a magnetic needle and a wire**

The separate parts of the magnetic needle at a rotation angle of $\theta$ are at different distances from the wire axis, so the magnitude of the magnetic field (4) exerted on them is different. Due to the inhomogeneity of the magnetic field round the wire the torque of the magnetic needle $M_I$ is obtained by the integration of the infinite small torques created by the different infinite small parts of the magnetic needle.

Each infinite small part of the needle of a length of **dl** is characterized by an infinite small dipole moment of **dm** having the same direction as that of the needle. The interaction of that part of the needle with the magnetic field of the wire leads to a torque of $dM_I$ [1]

$$dM_I = -\frac{\mu_0 \cdot I}{2 \cdot \pi} \cdot \frac{z \cdot \cos\theta}{\rho^2} \cdot dm. \qquad (6)$$

The torque of the whole needle, $M_I$, is:

$$M_I = -\frac{\mu_0 \cdot I}{2 \cdot \pi \cdot z} \cdot m \cdot \cos\theta \cdot \int_0^\Delta \frac{z^2}{\rho^2} \cdot \frac{dm}{m} = -\frac{\mu_0 \cdot I}{2 \cdot \pi \cdot z} \cdot m \cdot \cos\theta \cdot f(\delta) \qquad (7)$$

The integral on the right side of (7) is a dimensionless function $f(\delta)$ of the parameter $\delta$ (which is also dimensionless)

$$\delta = \frac{\Delta \cdot \sin\theta}{2 \cdot z}. \qquad (8)$$

The three-dimensional configuration in Figure 2 represents the torque having the same direction as that of the magnetic field of the wire. Obviously, this can be done by adjusting the north-south direction of the needle to the positive **x**-axis using the torque.

The functional dependency of $f(\delta)$ is determined by the shape of the needle.

Generally, the function of $f(\delta)$ is even and decreasing with an increase of $\delta$ due to the needle symmetry taking into account the pivot. At zero, the function of $f(\delta)$ has the value of 1.

For example, if the needle has a rectangular shape, the solution to the integral is:

$$f(\delta) = \frac{\text{arctg}(\delta)}{\delta} \cong 1 - \frac{\delta^2}{3} + \frac{\delta^4}{5} - \frac{\delta^6}{7} + \ldots \qquad (9)$$

The function of $f(\delta)$ is an important factor and it should be taken into consideration for comparable $\Delta$ and **z**. However, there is a great variety of magnetic needle shapes manufactured by different companies. Therefore, the dependency of $f(\delta)$ cannot be considered as universally valid and its theoretical values are a result of complex mathematical expressions. A tabulation of the function obtained experimentally should be used in such cases.

*2.3 A magnetic needle in the magnetic field of the Earth*

The Earth has its constant magnetic field. If the needle was placed only in the Earth's magnetic field, it would orient to the magnetic meridian of the geographical point, where the compass lies. That direction is to be nominated as north.

The geomagnetic field is a homogeneous one. The horizontal component of the induction of Earth's magnetic field orients the positive direction of **y**-axis (Fig. 2) and has a magnitude of $\mathbf{B_e}$. The interaction of the needle with it results in a torque of $\mathbf{M_e}$

$$\mathbf{M_e} = -m.B_e.\sin\theta. \qquad (10)$$

*2.4 A magnetic needle in Earth's magnetic fields and the wire*

We assume the magnetic needle has a moment of inertia, $\mathbf{J}$, towards its pivot. Taking onto consideration (7) and (10), the simultaneous action of both the Earth's magnetic field and the one created by the wire can be expressed by following equation of twisting motion:

$$\mathbf{J}.\frac{d^2\theta}{dt^2} = -m.B_e.\sin\theta - m.\frac{\mu_0.I}{2.\pi.z}.\cos\theta.f(\delta). \qquad (11)$$

The above equation has the following stationary solutions:

A) at zero current along the wire and applying (11), we obtain: $\theta = 0$; and the needle orients the Earth's magnetic field.

B) at non-zero current along the wire and applying (11), we obtain:

$$\frac{tg\theta}{f(\delta)} = -\frac{\mu_0}{2.\pi.z.B_e}.I = -\frac{I}{I_e}. \qquad (12)$$

where $\mathbf{I_e}$ is the equivalent to the current of Earth's magnetic field when **z** is given:

$$\mathbf{I_e} = \frac{2.\pi.z.B_e}{\mu_0}. \qquad (13)$$

Equation (12) expresses the dependency of the deflection angle of the needle on the wire current. Hence, this dependency can be used as a method for measuring the current magnitude along a wire. It could be called tangent-galvanometer and that is how the vertical current coils are called [4, 5].

The minus (-) sign in (12) means that when the current flows along the positive direction of **OY** axis the needle deviates on the right.

**3. Experimental Setup**

*3.1 A planar coil of wires*

Our next step was to devise and make a planar coil of wires with overall dimensions of 20 cm long and 30 cm wide to wind the wire in the middle section as many times as possible and create a bundle of windings there.

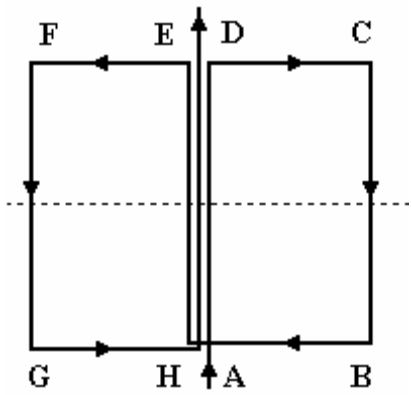 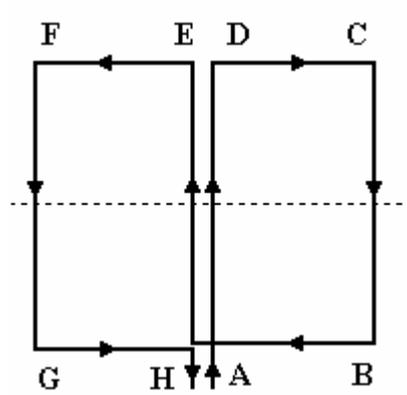

**Figure 3a.** A planar coil of uneven windings

**Figure 3b.** A planar coil of even windings

The magnetic field created by the side wires in the middle of the bundle of wires is compensated because areas AB and GH, BC and FG, CD and EF mutually neutralize their fields due to the symmetrical location of the areas and to the fact that the current flows along them symmetrically. Thus, the field created by the bundle of wires in the middle of the coil is what should be registered.

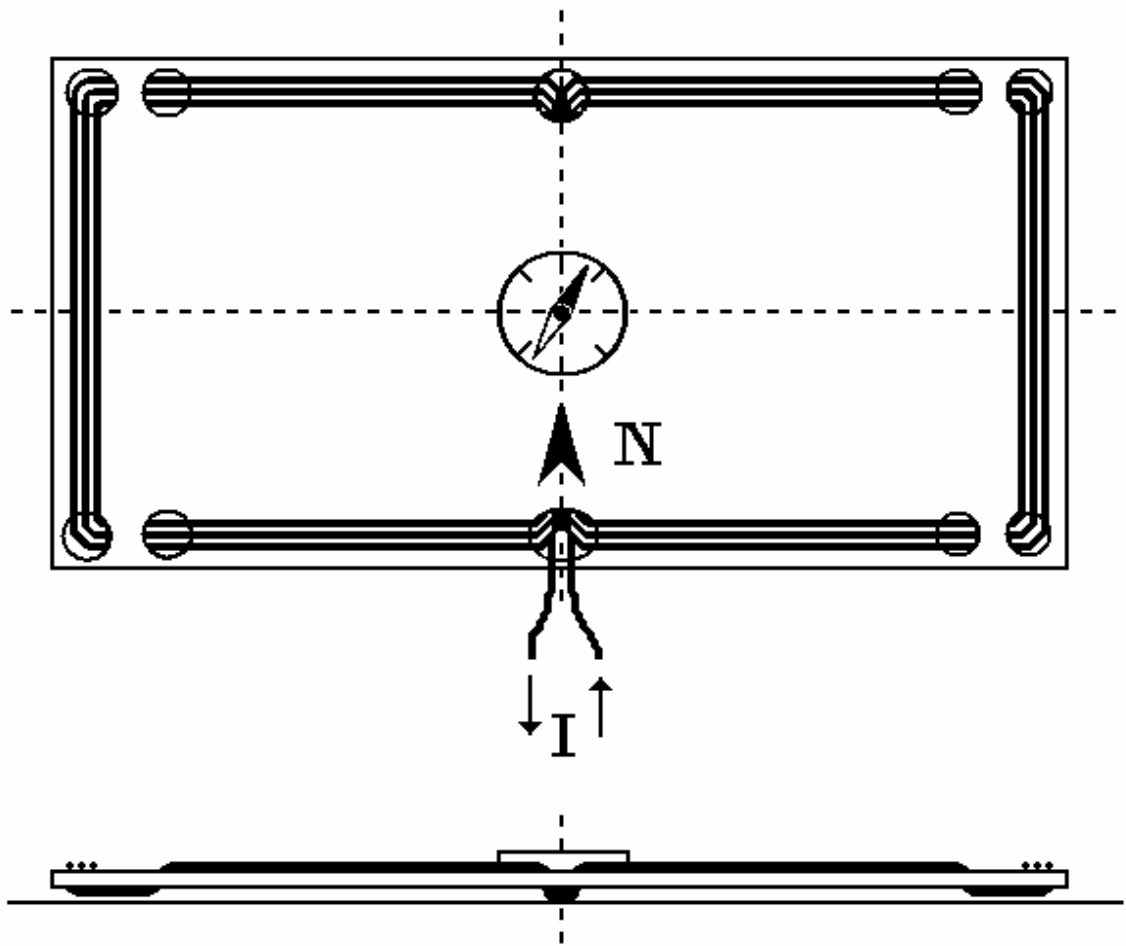

**Figure 4. Framework of the planar coil of wires used in the setup**

Figure 3a shows an uneven number of windings in the bundle while in figure 3b the number of windings in the bundle is even. In the setup we preferred the option with the even number of windings taking into consideration that the current input and output were located on one side of the plane framework.

We used a dielectric sheet (plastics sheet in our setup) and 8-mm holes were drilled into it. The location of the holes is presented in Figure 4.

The wire was wound onto the sheet while the holes were used to fasten the framework. The bundle of windings was located under the dielectric sheet. When at work, the planar coil was placed horizontally and the compass lay on the dielectric sheet in its geometrical centre. It was the sheet gauge that determined the minimum distance between the compass and the bundle windings.

### *3.2 The horizontal frame as a source of a magnetic field*

In the setup we used the planar coil with even number of windings. The total number of windings in the bundle was 24. A steady current was supplied by a DC power supply rectifier with an output voltage up to 15 V and a max current of 2 A. Resistance of 15 $\Omega$ / power of 20 W/ was connected in serial into the power supply and the planar coil as to be used as a ballast.

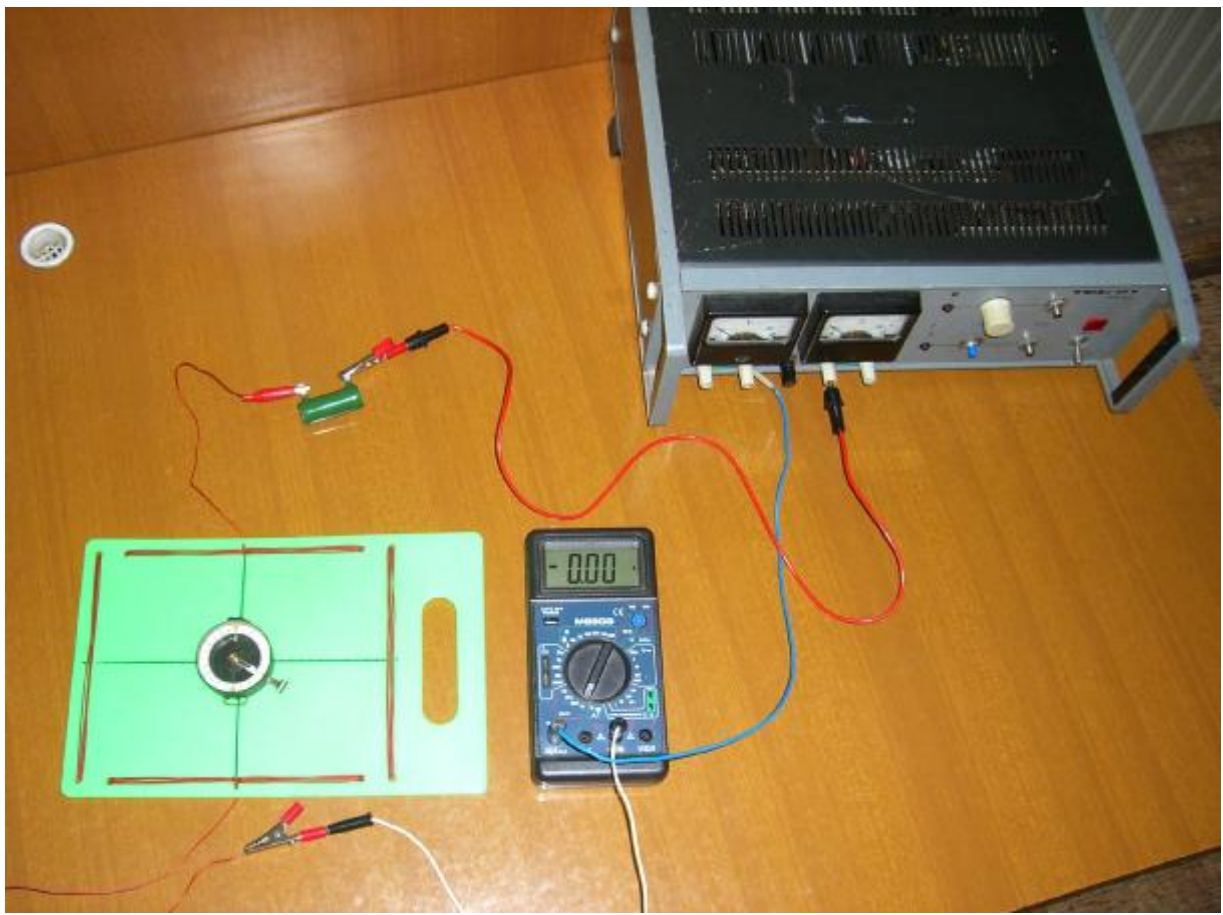

**Figure 5. The bunch windings oriented the north-south direction by the compass.**

At work the planar coil was put on horizontal table and the compass lay on the dielectric sheet in its geometrical centre (Figure 5). The bunch windings oriented the north-south direction determined by the compass as the current was switched off along the frame. The compass scale was set to zero when the coil was oriented that way and the current was switched off, the magnetic needle pointed north, i.e. $\theta = 0$.

Thus, the magnetic field created by the bunch of windings $\mathbf{B_I}$ had a direction perpendicular to the direction of the horizontal component of Earth's magnetic field, $\mathbf{B_e}$.

When the current was switched on along the coil, the magnetic needle deflected left or right at the angle of $\theta$ with regards to the current direction (See Figure 4). The angle of deflection on the left has a positive sign.

Current of **I** flowed along each winding. The range of change of **I** was from **-0.87 A** to **0.87 A** at a maximum deflection of $\theta = 84^0$.

Figure 6 represents the measurements of the deflection angle of $\theta$ as a multiplication of the current **I** along each wire by the number of windings, **N**.

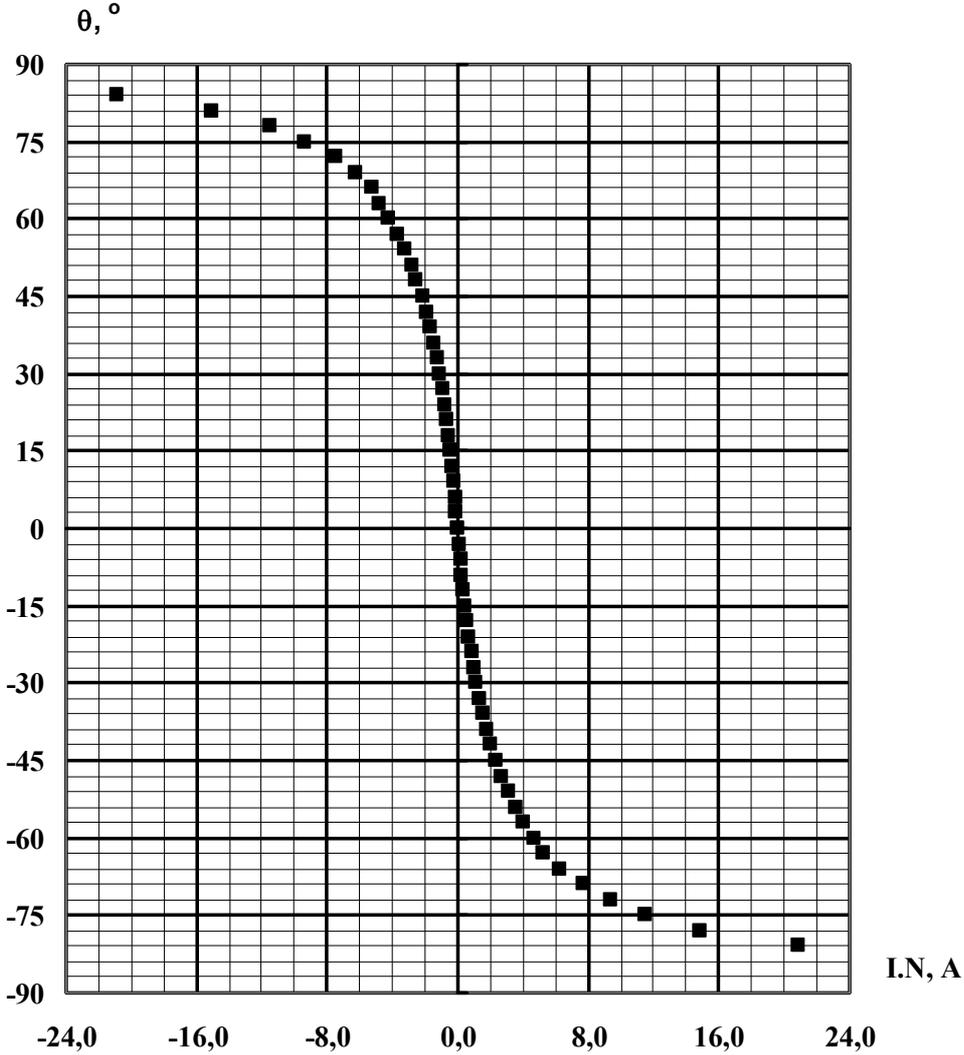

**Figure 6. Deflection angles of the magnetic needle as a multiplying function of the current I by the number of wires N**

# 4 Consideration of Results

## *4.1 The planar coil as a practical modification of Oersted experiment*

The planar coil framework shown in Figure 4 is one of the many examples which could be used when we do not want to use metal fastening elements or glue. Certainly, there are many other options as well.

The planar coil in the setup makes the needle deflect vigorously (more than 80 angular degrees) when the current flow is up to 1 A.

The magnetic needle can considerably deflect if a small battery of 1.5 A is directly connected to the coil as a voltage source (without any ballast resistance). Thus, Oersted experiment can be conducted in an easy and simple setup.

If the planar coil of wires has $L/z < 10$, the same results are obtained: the current carrying wire creates a magnetic field and the field direction is dependent on the current direction. However, precise quantitative measurements will show serious deviations from Biot – Savart law ranging from 10 to 15 %.

Figure 4 represents a planar coil of wires of $L/z > 20$. The magnetic field created by the bundle of windings and of accuracy better than 1% adjusted the magnetic field of a single straight infinite wire carrying a current of **I.N.** Consequently, it can be used as a modification of Oersted experiment.

## *4.2 The planar coil of wires as a tangent-galvanometer*

The presence of a source of magnetic field adjusting the magnetic field of a single straight infinite rectilinear current carrying wire can be used and be implemented in many other setups with practical application.

The abovementioned theoretical assumption (12) to use the planar coil of wires as a tangent-galvanometer could become reality if an accurate graduated curve was to be drawn to take into account nonlinearities.

## *4.3 Measurement of the horizontal component of Earth's magnetic field*

We would like to point out to the following procedure for measuring the horizontal component of Earth's magnetic field. It is based on the abovementioned theoretical analysis and on the possibility for planar coil of wires to be used as a tangent-galvanometer.

Firstly, we transform (12) by using (8) into:

$$-\frac{tg\theta}{I.N} = \frac{1}{I_e}.f\left(\frac{L.\sin\theta}{2.z}\right). \tag{14}$$

Then, the experimental results shown in Figure 6 are processed by **MS Excel**.

Figure 7 represents the left side of equation (14) as a function of $\sin\theta$, i.e. the dynamics of argument $\delta$ (8). The inhomogeneity of the magnetic field of the wire leads to the dynamic character of $f(\delta)$.

Next, the graphic dependency of the obtained experimental data is automatically approximated by using **Trendline MS Excel** function which conducts a polynomial approximation of the experimental points.

**Trendline MS Excel** automatically performs the following substitution (abscissa and ordinate in the graph):

$$y = -\frac{\text{tg}\theta}{I \cdot N}, \qquad (15)$$

$$x = \sin\theta. \qquad (16)$$

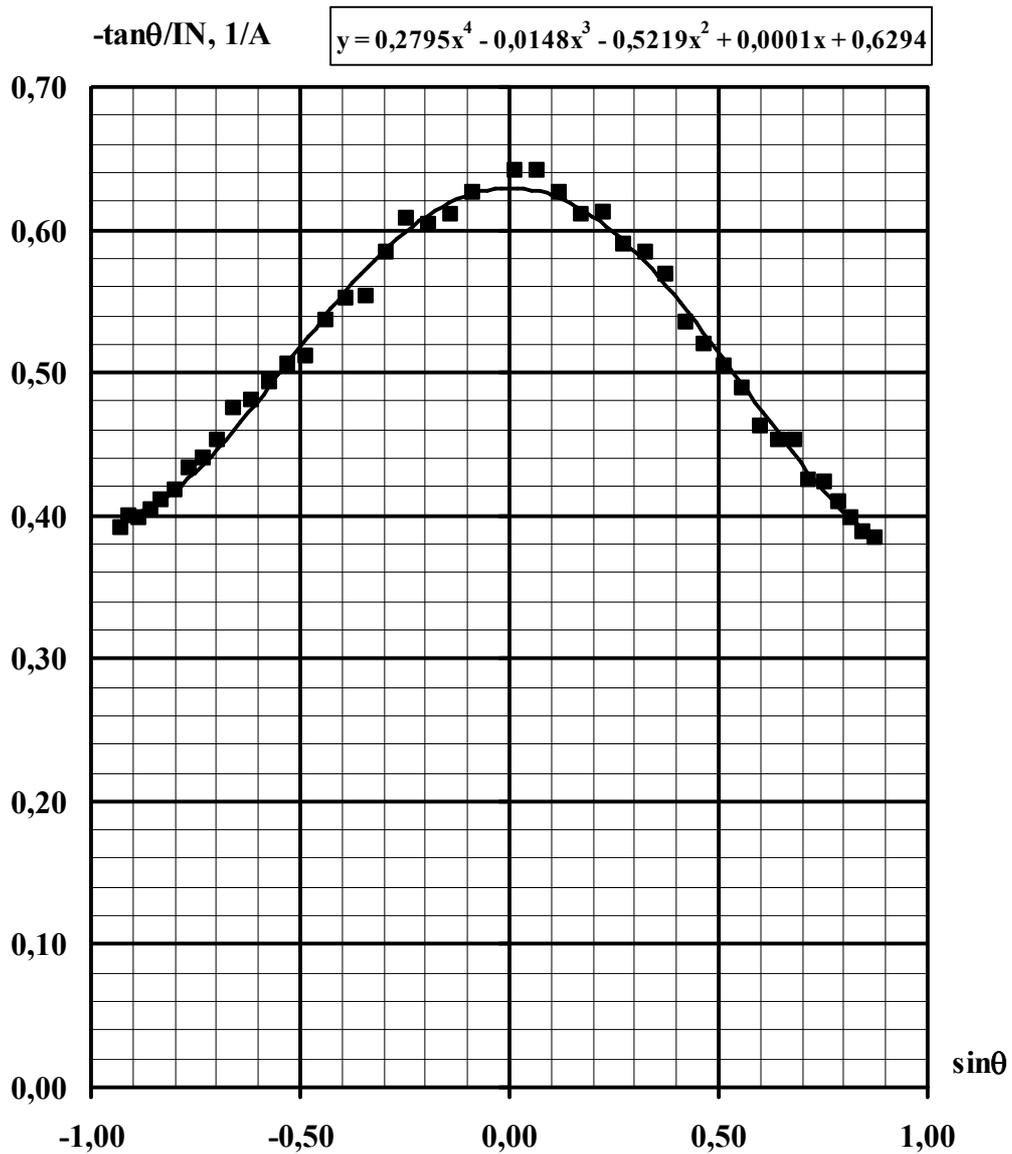

Figure 7. The dependency of the left side in (14) on $\sin\theta$.

At $\sin\theta = 0$, the value of the approximating function (shown in Figure 7) is as follows:

$$y(0) = 0.6294.$$

Taking into account $f(0) = 1$ and comparing (14) and (15), we obtain:

$$1/I_e = 0.63 \pm 0.02 \text{ A}^{-1} \tag{17}$$

The variations of each experimental points around the approximating function (valued at $\pm 0.02 \text{ A}^{-1}$) shown in Figure 6 are taken into account in (17). These variations are dependent on the measured accuracy of the deflection angle of the magnetic needle, $\theta$.

In the setup we measured the distance $z$, from the axis of the bundle of windings to the magnetic needle and it was: $z = 11.9 \text{ mm}$. The obtained $z$, the results from (17) together with (13) led to:

$$B_e = (26.7 \pm 0.8).10^{-6} \text{ T} \tag{18}$$

Hence, the magnitude of the horizontal component of Earth's magnetic field in the area of Sliven Engineering and Pedagogical Faculty was estimated.

## 4. Conclusion

Firstly, the proposed type of apparatus for Oersted experiment is very simple. The planar coil of wires makes the magnetic field deflect vigorously (more than 80 angular degrees) when current up to 1 A flows along it. Furthermore, a considerable deflection can be observed by using a small battery of 1.5 V as a power supply, connected with planar coil of wires without ballast.

Secondly, the theoretical analysis points out to dependency of the magnetic needle torque on the shape of the magnetic needle taking into account the inhomogeneity of the magnetic field around the current carrying wire. Furthermore, it outlines that the planar coil of wires can be used as a tangent-galvanometer.

Finally, a measuring procedure of the horizontal component of Earth's magnetic field is implied and implemented and its magnitude is estimated.